\documentstyle[12pt]{article}
\newread\epsffilein    
\newif\ifepsffileok    
\newif\ifepsfbbfound   
\newif\ifepsfverbose   
\newdimen\epsfxsize    
\newdimen\epsfysize    
\newdimen\epsftsize    
\newdimen\epsfrsize    
\newdimen\epsftmp      
\newdimen\pspoints     
\def\epsfbox#1{\global\def\epsfllx{72}\global\def\epsflly{72}%
   \global\def\epsfurx{540}\global\def\epsfury{720}%
   \def\lbracket{[}\def\testit{#1}\ifx\testit\lbracket
   \let\next=\epsfgetlitbb\else\let\next=\epsfnormal\fi\next{#1}}%
\def\epsfgetlitbb#1#2 #3 #4 #5]#6{\epsfgrab #2 #3 #4 #5 .\\%
   \epsfsetgraph{#6}}%
\def\epsfnormal#1{\epsfgetbb{#1}\epsfsetgraph{#1}}%
\def\epsfgetbb#1{%
\openin\epsffilein=#1
\ifeof\epsffilein\errmessage{I couldn't open #1, will ignore it}\else
   {\epsffileoktrue \chardef\other=12
    \def\do##1{\catcode`##1=\other}\dospecials \catcode`\ =10
    \loop
       \read\epsffilein to \epsffileline
       \ifeof\epsffilein\epsffileokfalse\else
          \expandafter\epsfaux\epsffileline:. \\%
       \fi
   \ifepsffileok\repeat
   \ifepsfbbfound\else
    \ifepsfverbose\message{No bounding box comment in #1; using defaults}\fi\fi
   }\closein\epsffilein\fi}%
\def\epsfsetgraph#1{%
   \epsfrsize=\epsfury\pspoints
   \advance\epsfrsize by-\epsflly\pspoints
   \epsftsize=\epsfurx\pspoints
   \advance\epsftsize by-\epsfllx\pspoints
   \epsfxsize\epsfsize\epsftsize\epsfrsize
   \ifnum\epsfxsize=0 \ifnum\epsfysize=0
      \epsfxsize=\epsftsize \epsfysize=\epsfrsize
     \else\epsftmp=\epsftsize \divide\epsftmp\epsfrsize
       \epsfxsize=\epsfysize \multiply\epsfxsize\epsftmp
       \multiply\epsftmp\epsfrsize \advance\epsftsize-\epsftmp
       \epsftmp=\epsfysize
       \loop \advance\epsftsize\epsftsize \divide\epsftmp 2
       \ifnum\epsftmp>0
          \ifnum\epsftsize<\epsfrsize\else
             \advance\epsftsize-\epsfrsize \advance\epsfxsize\epsftmp \fi
       \repeat
     \fi
   \else\epsftmp=\epsfrsize \divide\epsftmp\epsftsize
     \epsfysize=\epsfxsize \multiply\epsfysize\epsftmp   
     \multiply\epsftmp\epsftsize \advance\epsfrsize-\epsftmp
     \epsftmp=\epsfxsize
     \loop \advance\epsfrsize\epsfrsize \divide\epsftmp 2
     \ifnum\epsftmp>0
        \ifnum\epsfrsize<\epsftsize\else
           \advance\epsfrsize-\epsftsize \advance\epsfysize\epsftmp \fi
     \repeat     
   \fi
   \ifepsfverbose\message{#1: width=\the\epsfxsize, height=\the\epsfysize}\fi
   \epsftmp=10\epsfxsize \divide\epsftmp\pspoints
   \vbox to\epsfysize{\vfil\hbox to\epsfxsize{%
      \includegraphics{#1}%
      \hfil}}%
\epsfxsize=0pt\epsfysize=0pt}%

{\catcode`\%=12 \global\let\epsfpercent=
\long\def\epsfaux#1#2:#3\\{\ifx#1\epsfpercent
   \def\testit{#2}\ifx\testit\epsfbblit
      \epsfgrab #3 . . . \\%
      \epsffileokfalse
      \global\epsfbbfoundtrue
   \fi\else\ifx#1\par\else\epsffileokfalse\fi\fi}%
\def\epsfgrab #1 #2 #3 #4 #5\\{%
   \global\def\epsfllx{#1}\ifx\epsfllx\empty
      \epsfgrab #2 #3 #4 #5 .\\\else
   \global\def\epsflly{#2}%
   \global\def\epsfurx{#3}\global\def\epsfury{#4}\fi}%
\def\epsfsize#1#2{\epsfxsize}
\let\epsffile=\epsfbox

\newlength{\dinwidth}
\newlength{\dinmargin}
\setlength{\dinwidth}{21.0cm}
\textheight24.2cm \textwidth17.1cm
\setlength{\dinmargin}{\dinwidth}
\addtolength{\dinmargin}{-\textwidth}
\setlength{\dinmargin}{0.5\dinmargin}
\oddsidemargin -1.0in
\addtolength{\oddsidemargin}{\dinmargin}
 
\setlength{\evensidemargin}{\oddsidemargin}
\setlength{\marginparwidth}{0.9\dinmargin}
\marginparsep 8pt \marginparpush 5pt
\topmargin -42pt
\headheight 12pt
\headsep 30pt 
\parskip 3mm plus 2mm minus 2mm
\parindent 5mm

\newcommand{\bfig}{\begin{figure}}
\newcommand{\efig}{\end{figure}}
\newcommand{\bcen}{\begin{center}}
\newcommand{\ecen}{\end{center}}
\newcommand{\beq}{\begin{equation}}
\newcommand{\eeq}{\end{equation}}
\newcommand{\btabu}{\begin{tabular}}
\newcommand{\etabu}{\end{tabular}}
\newcommand{\btabl}{\begin{table}}
\newcommand{\etabl}{\end{table}}

\def\lsim{\mathrel{\rlap{\lower4pt\hbox{\hskip1pt$\sim$}}
    \raise1pt\hbox{$<$}}}         
\def\gsim{\mathrel{\rlap{\lower4pt\hbox{\hskip1pt$\sim$}}
    \raise1pt\hbox{$>$}}}         

\def\lsim{\mathrel{\rlap{\lower4pt\hbox{\hskip1pt$\sim$}}
    \raise1pt\hbox{$<$}}}         
\def\gsim{\mathrel{\rlap{\lower4pt\hbox{\hskip1pt$\sim$}}
    \raise1pt\hbox{$>$}}}         

\begin{document}


\title{DIPSI: a Monte Carlo generator for elastic vector meson
production in charged lepton-proton scattering
}

\author{M.~Arneodo$^{a}$, L.~Lamberti$^{b}$ and M.~Ryskin$^{c}$}

\date{
{\footnotesize$^{a}$ Universit\`a di Torino,via Giuria,1, I-10125 Torino, Italy and\\
Universit\`a della Calabria and INFN, I-87036 Arcavacata di Rende (Cs), Italy\\
ARNEODO@VXDESY.DESY.DE\\
$^{b}$ Universit\`a di Torino and INFN, via Giuria,1, I-10125 
Torino, Italy,\\
now at Imperial College, Blackett Laboratory, Prince Consort Rd.,
London SW7~2BZ, United Kingdom\\
LAMBL@VXDESY.DESY.DE\\
$^{c}$ Petersburg Nuclear Physics Institute, 188350 
Gatchina, Russia\\
RYSKIN@THD.PNPI.SPB.RU}
}
\markright{.}

\maketitle

\begin{abstract}
\noindent
We present a Monte Carlo generator for exclusive vector meson
production in charged lepton-proton interactions,
$l+p \rightarrow l+p+V$, based on a QCD leading logarithm
model calculation according to which the cross section for this
process is proportional to the square of the gluon momentum density in 
the proton. The generator can be used for both 
fixed target and collider kinematics.
Each event is assigned a weight equal to its cross
section, so that the independent variables can be generated according to
convenient distributions, not necessarily identical to the physics ones.
\end{abstract}

\section{Introduction}

In this paper we present a Monte Carlo generator for
exclusive production of vector mesons in the reaction
\begin{eqnarray}
e (\mu) + p \rightarrow e (\mu) + p + V,
\label{reaction}
\end{eqnarray}

\noindent
where an electron ($e$) or a muon ($\mu$) interacts with a proton
($p$), producing, in the final state, a vector meson
$V$ ($\rho^0$, $\omega$, $\phi$, $J/\psi$...).
This reaction is also referred to as elastic vector meson production.

\begin{figure}
\vspace{-1.0cm}
\begin{center}
\leavevmode
\hbox{%
\hspace*{-0.8cm}
\epsfxsize = 10.0cm
\epsffile{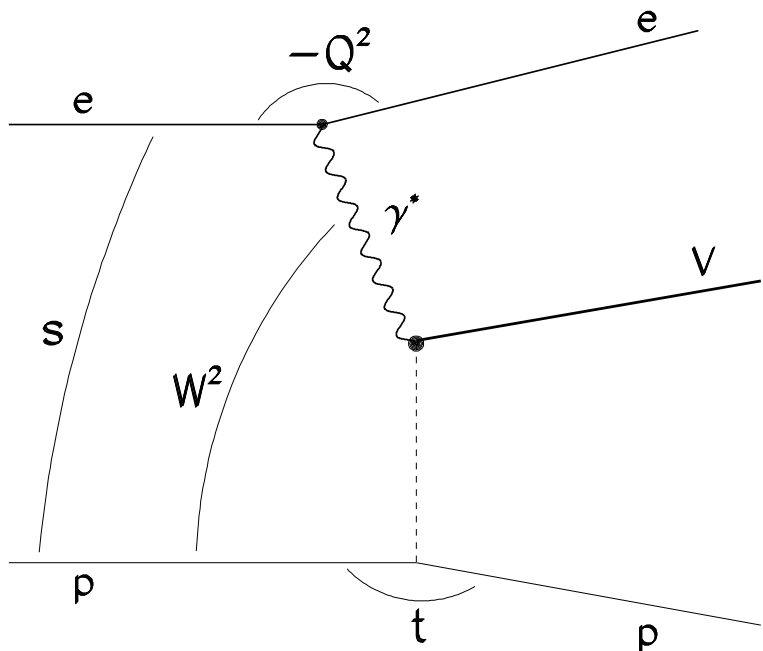}}
\end{center}
\vspace{-1cm}
\caption{Elastic vector meson production.}
\label{vel}
\end{figure}

Figure \ref{vel} shows the diagram of the process; tables
\ref{variables1} and \ref{variables2} define the relevant kinematic variables. The incoming
lepton radiates a virtual photon of mass $-Q^2$ that collides with the
target proton, turning into a vector meson; the proton emerges intact
from the interaction. Since the vector meson has the same quantum
numbers as the photon, the object exchanged between the photon and the
proton, a pomeron, carries the quantum numbers of the vacuum.

\begin{figure}
\vspace{-1.0cm}
\begin{center}
\leavevmode
\hbox{%
\hspace*{-0.8cm}
\epsfxsize = 13.0cm
\epsffile{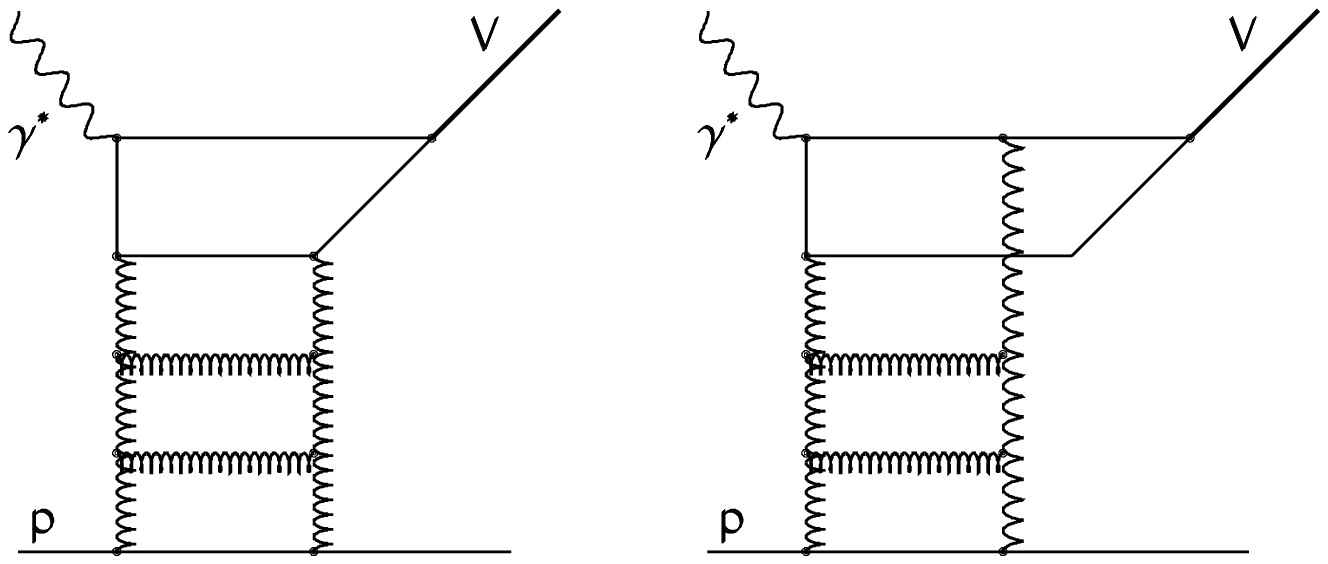}}
\end{center}
\vspace{-1cm}
\caption{Elastic vector meson production in LLA.}
\label{rhodiag}
\end{figure}


{\footnotesize
\begin{table}
\begin{center}
\begin{tabular}{|l|l|} \hline
$k$                  & four-momentum of the incident lepton\\
$k^{\prime}$         & four-momentum of the scattered lepton\\
$p$                  & four-momentum of the incident/target proton \\
$p^{\prime}$         & four-momentum of the scattered proton \\
$v$                  & four-momentum of the produced vector meson\\
$m_V$                & mass of the vector meson\\
$m_p$                & mass of the target proton\\
$q=k-k^{\prime}$     & four-momentum of the virtual photon\\
$-Q^2=q^2$           & invariant mass squared of the virtual photon\\
$y=p\cdot q/p\cdot k$& fraction of the beam energy carried by the photon \\
                     & (in the proton rest frame)\\
$\nu=p\cdot q/m_p$   & photon energy (in the proton 
rest frame)\\
$W=\sqrt{(q+p)^2}$   & photon-proton centre of mass energy \\
$\sqrt{s}=\sqrt{(k+p)^2}$& electron-proton centre of mass energy \\
\hline
\end{tabular}
\end{center}
\caption{Definition of the relevant kinematic variables, part 1.}
\label{variables1}
\end{table}
}

{\footnotesize
\begin{table}
\begin{center}
\begin{tabular}{|l|l|} \hline
$t=(p-p^{\prime})^2$ & four-momentum squared exchanged at the proton vertex\\
$p_t$                & transverse momentum of the vector meson with respect to\\
                     & the virtual photon direction\\
$z=p\cdot v/p\cdot q$& fraction of the photon energy carried by the vector meson \\
                     & (in the proton rest frame)\\
$x_l = |\mbox{{\bf p}}^{\prime}|/|\mbox{{\bf p}}|$ & ratio between outgoing and 
incoming proton three-momenta \\
                     & in the laboratory frame\\
$p_t^p$              & tranverse momentum of the proton in laboratory frame\\
$\bar x = \frac {Q^2 +m_V^2 +p_t^2}{W^2}$
                     & fraction of the proton's momentum carried by the two-gluon \\
                     & system in the model of ref.~\protect\cite{misha}\\
$\bar q^2 = \frac {Q^2 +m_V^2 +p_t^2}{4}$
                     & scale at which the gluon density is probed in the model of ref.~\protect\cite{misha}\\
$\vartheta$          & polar angle of the positive decay particle (e.g. $\pi^+$ in the $\rho^0$ case)\\
                     & with respect to the outgoing $p$ direction in the $V$ rest frame\\
                     & (for two-body decays)\\
$\varphi$            & angle between the decay plane and the meson production\\
                     & plane, which contains the virtual photon and the meson, in the\\
                     & $V$ rest frame (for two-body decays)\\
$\Phi$               & angle between the lepton scattering plane and the vector\\
                     & meson production plane (in the proton rest frame)\\
$\psi=\varphi-\Phi$  & \\
\hline
\end{tabular}
\end{center}
\caption{Definition of the relevant kinematic variables, part 2.
}
\label{variables2}
\end{table}
}

Pomerons were originally introduced to describe diffractive
interactions be\-tween ha\-drons. Elastic vector meson production 
at small $Q^2$ can indeed be thought of as a hadronic process 
in the Vector Meson Dominance (VMD)~\cite{saku} framework, in 
which the photon is assumed to fluctuate into a vector meson 
before interacting with the proton. Reaction (\ref{reaction}) 
may thus allow the investigation of the properties of the 
pomeron, as well as of the hadronic features of the photon.

Elastic vector meson production has been extensively studied in 
fixed target experiments in the photoproduction regime
($Q^2=0$)~\cite{bible}-\cite{omega} and at larger values of
$Q^2$~\cite{chio}-\cite{nmc},
for photon-proton centre of mass energies $W \lsim 20$~GeV.
With the advent of the electron-proton collider HERA,
reaction~(\ref{reaction}) has also been investigated at
values of $W \approx 100$~GeV, both for
$Q^2~\approx~0$~\cite{zeus_rho}-\cite{zeus_psi} and for non-zero
virtualities~\cite{zeus_rhohiq2}-\cite{zeus_phihiq2}.

From the theoretical point of view, several
models~\cite{dl}-\cite{kolya} offer a description of elastic vector
meson production; all assume pomeron exchange, with the pomeron
described as a gluon pair. Individual models differ in the way the
gluons are treated, ranging from non-perturbative approaches to
perturbative ones.

The generator we wrote, hereafter referred to as DIPSI, is based 
on the model of ref.~\cite{misha}\footnote{The results of
ref.~\cite{brodsky} lead to formulae with the same structure.}.
This model assumes that the exchanged virtual
photon fluctuates into a $q\bar{q}$ pair which then interacts with a
gluon ladder emitted by the incident proton (fig.~\ref{rhodiag}).
The parameters of the
model are the strong coupling constant $\alpha_s$, the two-gluon form
factor of the proton and the gluon momentum density in the proton. Once these
are chosen, the dependences on $W$ and $t$, where $t$ is the
four-momentum squared transferred at the proton vertex, are fixed.

DIPSI has been extensively used within the ZEUS experiment --~operating 
at the electron-proton collider HERA at DESY~-- in
particular for the determination of the detector acceptance for
$\rho^0$~\cite{zeus_rho,tesi_luc}, $\phi$~\cite{zeus_phi} 
and $J/\psi$ 
photoproduction~\cite{zeus_psi}
as well as for $\rho^0$~\cite{zeus_rhohiq2} and
$\phi$~\cite{zeus_phihiq2} production at large $Q^2$.

The paper is organised as follows. In section~\ref{model} we give
a short description of the model and in section~\ref{program} we present
the program itself. The appendices contain a list of the files provided, 
a description of the input control cards as well as a list of 
the routines and functions used in the program.

\section{The model}
\label{model}

The model~\cite{misha} is based on a perturbative QCD (pQCD) 
calculation of the $\gamma\to V$
process and assumes a non-relativistic wavefunction for the vector meson
$V$. In the Born approximation, at small values of the parton fractional 
momentum $x$, the amplitude for the 
reaction $\gamma p \rightarrow V p$ is given 
by the sum of those of the graphs shown in fig.~\ref{rhodiag}
(without the ``horizontal" gluons which form the rungs of the ladder),
where the two $t$-channel gluons play the role of the pomeron 
(in the Born Low-Nussinov approximation~\cite{low-nussinov}). 
It is convenient to write the 
contribution of the upper quark loop in terms of the electronic width
$\Gamma^V_{\ell\ell}=\Gamma$ of the vector meson. In this way a
large part of the $O(\alpha_s)$ corrections is absorbed in the
experimental value of $\Gamma$; here $\alpha_s$ is the strong coupling 
constant. To take into account 
logarithmic corrections which contain $(\alpha_s\ln q^2)^n$ or 
$(\alpha_s\ln 1/x)^m$ terms (here $q^2$ denotes the scale relevant for 
the process), one has to sum up a set of
more complicated Feynman diagrams. In the axial gauge,
these diagrams have the ladder form (i.e. include all gluons 
shown in fig. \ref{rhodiag}). At small $t$ the lower
part (that below the fermion loop) of the graphs of
fig.~\ref{rhodiag} may be considered as the gluon momentum density in the 
proton $xg(x,q^2)$. Indeed, 
for $t=0$ this lower part of the graph contains exactly the same 
Feynman diagrams (calculated for the same kinematics) as those involved 
in the definition of $xg(x,q^2)$ in the Leading Logarithm Approximation 
(LLA) QCD treatment of deep inelastic scattering.

The arguments $x$ and $q^2$ of $xg(x,q^2)$ take the values 
$\bar x= (Q^2 +m_V^2+p^2_t)/W^2$  and $\bar q^2 =(Q^2 +m_V^2 +p^2_t)/4$, 
respectively.
The value of $\bar x$ is larger than that of the Bjorken variable 
$x_{Bj} \simeq Q^2/W^2$, as one has to transfer additional longitudinal 
momentum along the pomeron (i.e. the $t$-channel gluons) to put the 
vector meson on mass shell.

The typical virtuality $\bar q$ was chosen equal to $\frac{1}{2}
\sqrt{Q^2 +m_V^2 +p^2_t}$ as the $t$-channel gluons interact with
one quark only, which, in the non-relativistic approximation used in the
model, carries one half of the external 
momenta (i.e. those of the photon and of vector meson). 

At large energy the value of $\bar x$ is small and for zero angle
production, when $t=t_{\min} \simeq -\bar x^2 m^2_N \to0$,  
we can write the cross section in terms of $\Gamma$
and $g(\bar x, \bar q^2)$. An explicit calculation gives: 

\begin{eqnarray} 
\frac{d\sigma^T(\gamma p \to V p)}{dt} & = &  
\frac{\alpha_s^2(\bar q^2) \Gamma m_V^3}{3 \alpha} \pi^3 \times \nonumber \\ 
& & \left[\bar xg(\bar x, \bar q^2) 
\frac{f(\bar q^2,p_t^2)}{2 \bar q^2(2\bar q^2-p_t^2)\ln{(8\bar q^2/p_0^2)}}\right]^2
\left[F_N^{2G}(t)\right]^2  \eta^2,
\label{form} 
\end{eqnarray} 

\noindent
where $\alpha=1/137$ is the electromagnetic coupling constant and the function 
$f(\bar q^2,p_t^2)$ is defined as follows:

\begin{eqnarray} 
f(\bar q^2,p_t^2)=\ln{\left[\frac
{4\bar q^2-p_t^2+p_0^2} {p_t^2+p_0^2}\right]}~{\mbox{for $p_t^2 \le p_0^2$}},\\
f(\bar q^2,p_t^2)=
\ln{\left[\frac{p_t^2+p_0^2} 
{4\bar q^2-p_t^2+p_0^2}\frac{4\bar q^4}{p_t^4}\right]}~{\mbox{for $p_t^2 > p_0^2$}},
\label{akin} 
\end{eqnarray} 

\noindent
with the infrared cutoff parameter $p_0^2=0.5$~GeV$^2$. 

Part of the $t$ (i.e. $p^2_t$) dependence of the amplitude is given by 
the bottom part of the graphs of fig.~\ref{rhodiag}, i.e. by 
the spatial distribution of the 
colour charge inside the proton. We parametrise this distribution by 
means of the two-gluon form factor $F^{2G}_N(t)$ (with $F^{2G}_N(0)=1)$. 
This form factor cannot be calculated within pQCD and should be 
measured experimentally. 
The simplest hypothesis is that the function $F^{2G}_N(t)$ is close
to the electromagnetic form factor and can thus be written, for 
instance, as
$[F^{2G}_N(t)]^2=e^{Bt}$, with  $B \approx 5$~GeV$^{-2}$
\footnote{However estimates based upon QCD sum 
rules~\protect\cite{braun} give somewhat smaller values of $B$.}; 
equivalently the dipole approximation can be used. This hypothesis
is implemented in the program.
Unfortunately, within the LLA there is  no shrinkage of the
diffractive cone. Therefore the slope $B$ does not depend on energy.
This is a shortcoming of the model, and to imitate the effect of the
diffractive cone shrinkage one should choose different
values of $B$ for different energies.

The remaining part of the $t$-dependence is contained in the factor 
$\{f/[\bar q^2 (2\bar q^2-p_t^2) \ln{(8\bar q^2/p_0^2)}]\}^2$, which 
comes from the upper fermion 
loop in fig.~\ref{rhodiag}, as well as in the $\bar x$ and $\bar q^2$ 
dependence of the gluon distribution.

As mentioned above, a non-relativistic meson wavefunction is assumed.
It is possible however to allow for relativistic effects 
by shifting the overall normalisation by a factor $\eta^2$. 
For $\rho^0$ mesons $\eta \simeq 1.8$~\cite{relativistic}, while it is 
expected to be close to unity for $J/\psi$ mesons.

There are a few restrictions on the applicability of the model.
In the first place, to be sure that $|t_{\min}|<\Lambda^2_{QCD}$ (or 
$|t_{\min}|< 1/R^2_N$, with $R_N$ the nucleon radius), only 
the region $\bar x<0.1$ should be considered.

Furthermore, strictly speaking, one should use the pQCD result only at 
sufficiently large values of $\bar q^2$, say $\bar q^2 >2$ GeV$^2$. 
Nevertheless, the program can be used also at smaller values of 
$\bar q^2$, e.g. for $\rho^0$ photoproduction, for which
$\bar q^2\simeq 0.15$ GeV$^2$. This is the reason for 
modifying formula (6) of ref.~\cite{misha} into~(\ref{form})
by including an infrared cutoff.
There are no singularities for $\bar q^2
>\Lambda^2_{QCD}$; however the applicability of pQCD  for these values 
of $\bar q^2$ is questionable. 
Parametrisations can be found for 
$\bar x g(\bar x, \bar q^2)$, with which the experimental results on $\rho^0$ 
photoproduction are reproduced~\cite{zeus_rho,tesi_luc}; in this region 
however it is not clear whether the function 
$\bar x g(\bar x, \bar q^2)$ can still be 
interpreted as the gluon momentum density in the proton.

We have so far discussed only the case of  transversely polarised photons.
The exchange of a pQCD (two-gluon) pomeron conserves 
$s$-channel helicity~\footnote{There was a mistake in
sect.~3 of ref.~\cite{misha}, where the helicity
flip amplitude $A_{1,0}$ was indicated to be non-zero.
One of us (M.R.) would like to thank D. Kr\"ucker for pointing this out.}
and,  if the vector meson wavefunction has the non-relativistic form,
the longitudinal cross section is given by $\sigma^L =
(Q^2/m^2_V)\sigma^T$. In the program both the $\sigma^L$ and
$\sigma^T$ terms are taken into account with the corresponding
angular distributions for the vector meson decay products.
As an example, in the case of two-body decays, the cross section is 
written as:

\begin{eqnarray}
\frac{d^2\sigma^T}{d\cos{\vartheta} d\bar\psi}=\frac{1}{4\pi} \sigma^T k_T(\cos{\vartheta},\bar \psi),\\
\frac{d^2\sigma^L}{d\cos{\vartheta} d\bar\psi}=\frac{1}{4\pi} \sigma^L k_L(\cos{\vartheta},\bar \psi),
 \label{angle_t_l}
\end{eqnarray}
\noindent 
where $\vartheta$ and $\psi$ are defined in table \ref{variables2} and
$\bar \psi=\psi+\delta$, with $\delta$ the angle that the photon 
polarisation vector forms with the normal to
lepton scattering plane; this angle is taken to be zero with probability 
$(2-y)^2/Ny^2$ and $90^{\circ}$ with probability $1/N$, where 
$N=1+(2-y)^2/y^2$. 

The angle $\psi$ is the difference of 
 the azimuthal angles $\varphi$ and $\Phi$, which are defined in two 
different frames. These frames however differ by a boost which is very 
nearly (to an accuracy better than $(Q^2+p_t^2)/s$) perpendicular to 
the quantisation axes of both frames, thus hardly affecting the 
azimuthal distributions.

For decays to spin-1/2 particles (e.g. $J/\psi \rightarrow e^+e^-$), 
one has $k_T(\cos{\vartheta,\bar \psi})=3/2(1-\sin^2{\vartheta}\cos^2{\bar\psi})$ 
and
$k_L(\cos{\vartheta})=3/2\sin^2{\vartheta}$. For decays to 
spin-0 particles
(e.g. $\rho^0 \rightarrow \pi^+\pi^-$), the angular distributions become
$k_T(\cos{\vartheta},\bar\psi)=3\sin^2{\vartheta}\cos^2{\bar\psi}$ and
$k_L(\cos{\vartheta})=3(1-\sin^2{\vartheta})$,
respectively.

In the case of the three-body decay $V \rightarrow \pi^+ \pi^- 
\pi^0$, the angular distributions are given by 
$k_T(\cos{\vartheta},\bar \psi) \propto \sin^2{\vartheta}\cos^2{\bar\psi}$ and
$k_L(\cos{\vartheta}) \propto (1-\sin^2{\vartheta})$,
where now $\vartheta$ and $\bar\psi$ are defined as in the two-body decay case but
for the normal to the decay plane rather than for the $\pi^+$ direction. 

For all other decay modes, the function $k$ is taken as constant.

The $\gamma p$ cross section is related to the $ep$ (or $\mu p$) cross 
section as follows:

\begin{eqnarray} 
\frac{d^2\sigma(ep \rightarrow epV)}{dy dQ^2} = 
\Gamma_T \sigma^T(\gamma p \rightarrow V p) + 
\Gamma_L \sigma^L(\gamma p \rightarrow V p),
\label{ep} 
\end{eqnarray} 

\noindent 
where $\Gamma_T$ and $\Gamma_L$ are the fluxes of transversely and 
longitudinally polarised virtual photons, respectively. Their explicit 
expression is:

\begin{eqnarray}
\Gamma_T= \frac{\alpha}{2\pi Q^2}\left[
   \frac{1+(1-y)^2}{y} - \frac{2(1-y)}{y}\frac{Q_{\min}^2}{Q^2}\right],\\
\Gamma_L= \frac{\alpha}{2\pi Q^2} \frac{2(1-y)}{y},
 \label{flux}
\end{eqnarray}
\noindent 
where $Q^2_{\min}= M_e^2 \frac{y^2}{1-y}$ is the minimum value of $Q^2$ 
kinematically allowed.

\section{The program}
\label{program}

DIPSI generates events according to reaction~(\ref{reaction}) in which
$V$ can be, at present:

\begin{itemize}
\item $\rho^0 \rightarrow \pi^+\pi^-$,
\item $\omega \rightarrow \pi^+\pi^-\pi^0$ or $\pi^+\pi^-$,
\item $\phi \rightarrow K^+K^-$, $K^0_L K^0_S$ or $\pi^+\pi^-\pi^0$,
\item $\rho(1450) \rightarrow \pi^+\pi^-$, $\pi^+\pi^- \rho^0$ or 
$\pi^0\pi^0 \rho^0$, 
\item $\rho(1700) \rightarrow \pi^+\pi^-$, $\pi^+\pi^- \rho^0$ or
$\pi^0\pi^0 \rho^0$, 
\item $J/\psi \rightarrow e^+e^-$, $\mu^+\mu^-$ or $\pi^+\pi^-\pi^0$,
\item $\psi^\prime (3600) \rightarrow e^+e^-$, $\mu^+\mu^-$,
                                      $\pi^+\pi^- J/\psi$
or $\pi^0\pi^0 J/\psi$, 
\item $\Upsilon \rightarrow e^+e^-$, $\mu^+\mu^-$ or $\pi^+\pi^-\pi^0$.
\end{itemize}

The program is steered by control cards, in which the user should
fix, for each run, some parameters and choose among a few options:

\begin{itemize}
\item lepton beam type (electron, muon);
\item beam momenta (for fixed target kinematics the proton beam 
momentum has to be set to 0; the lepton beam momentum must
always be negative);
\item generation limits for $Q^2$, $y$ and $p_t^2$ and type of
generation for each of these variables (as described in
section~\ref{weights});
\item type of meson produced, decay mode and mass spectrum;
\item value of $\alpha_s$, gluon density and proton form factor;
\item amount of output data.
\end{itemize}

Before discussing the structure of the program and its features, we
describe the method of weighted events on which the generator is based.

\subsection{The method of weighted events}
\label{weights}

Each event generated with DIPSI is assigned a weight equal to its cross
section. This makes it possible to decouple the generation of the independent
variables from their expected distribution.

As an example,
let us consider the $p_t^2$ distribution.
The model predicts an approximately exponential distribution of the type
$A\exp{(-b p_t^2)}$; suppose that for a given vector meson $b=10$
GeV$^{-2}$. If events are generated according to
such a distribution it may take considerable computer time to 
substantially populate the high $p_t^2$ region.
In order to study the high $p_t^2$ tail, one can instead choose to
generate with, say, $b=3$~GeV$^{-2}$.
To obtain a physically meaningful $p_t^2$ distribution one has then to
weight each event with its cross section and with a factor, that we will
name ``phase space factor'', which takes into account the way in which the
variable has been generated.

Assume, for the sake of simplicity, that the model had only one
independent variable, $\xi$. Suppose that, given a random number $R_i$
between zero and unity, sampled
from a uniform distribution, one obtains a value $\xi_i$ for $\xi$
using some algorithm $\xi=\xi(R)$ which is able to populate
adequately the region of $\xi$ under study.
Based on the model, the value of $d\sigma/d\xi_i$, the differential
cross section $d\sigma/d\xi$ for $\xi=\xi_i$, can then be computed. The
weight of the event is given by
\begin{eqnarray}
\sigma_i=\frac{d\sigma}{d\xi_i} \frac{d\xi}{dR_i},
\label{weight}
\end{eqnarray}

\noindent
where the phase space factor $d\xi/dR_i$ is the value of $d\xi/dR$
computed for $R=R_i$. The phase space factor would be unity if
$\xi$ were sampled from a uniform distribution in the range 0-1.

The cross section for the process, integrated over the generation range
of $\xi$, is thus
\begin{eqnarray}
\sigma=\frac{1}{N_{tot}} \sum_{i=1}^{N_{tot}} \sigma_i,
\label{sigma}
\end{eqnarray}

\noindent
where $N_{tot}$ is the total number of generated events. The statistical 
uncertainty on $\sigma$ is given by 

\begin{eqnarray}
\Delta\sigma=\frac{1}{N_{tot}}\sqrt{ \sum_{i=1}^{N_{tot}} \sigma_i^2}.
\label{uncertainty}
\end{eqnarray}

Likewise, the differential cross section $d\sigma/d\xi$ for $\xi=\xi_i$
is obtained as
\begin{eqnarray}
\frac{d\sigma}{d\xi}=\frac{1}{\Delta \xi} \sum_{bin} \sigma_i/N_{tot}
=\frac{1}{\Delta \xi N_{tot}} \sum_{bin} \sigma_i,
\label{sigma_diff}
\end{eqnarray}

\noindent
where the sum is extended to the events in a narrow bin of width 
$\Delta \xi$ centred around $\xi_i$.

\subsection{The structure of the program}

The code consists of three main routines: DIPSINI, which controls the
initialisation phase; DIPSIGEN,  for
the actual event generation; DIPSOUT, which terminates the program.

\subsubsection{Initialisation}
In DIPSINI the relevant variables are initialised, the random number 
seed and the external
input cards are read in and the histograms and n-tuples are booked.
The input quantities read from the control cards are described 
in appendix~\ref{control_cards}.

\subsubsection{Event generation}
For each event the following independent kinematic variables are
generated (subroutine DIPSIGEN):
$y$, $Q^2$, the mass of the vector meson $m_V$, $p_t^2$ and
the azimuthal angles of the scattered lepton with respect to the
incoming beam direction and of the vector meson with respect to the virtual
photon direction.
The four-momenta of the scattered lepton, of the
vector meson and of the scattered proton are then fixed by energy-momentum
conservation.

Various options, detailed in appendix~\ref{control_cards}, are available
for generating $y$ (flat, $1/y$), $Q^2$ (flat, $1/Q^2$, $1/Q^4$) and
$m_V$ (flat, Breit-Wigner, S\"oding~\cite{soeding,zeus_rho}, relativistic 
p-wave Breit-Wigner). The variable $p_t^2$ is sampled from an exponential 
distribution of which the user can choose the slope.

Once the kinematics of the event has been generated, the cross section
is evaluated as discussed in sect.~\ref{model} by subroutine JPRYSKIN,
called by DIPSIGEN. Immediately after calling JPRYSKIN, DIPSIGEN 
evaluates the weight by multiplying the cross section by the phase space 
factors.
In addition to the weight for the process $ep\rightarrow epV$, that
for the reaction $\gamma^* p \rightarrow V p$ is also computed.

Finally subroutines JDK (for two-body decays) and TREDK (for three-body 
decays), also called by DIPSIGEN, let the vector meson decay. At this 
point the weight is modified so as to take 
into account the dependence of the cross section on the decay angular 
variables; these variables are evaluated by the routines DPHELI and 
HELOMEGA for two-body and three-body decays, respectively.

\subsubsection{Termination}

Subroutine DIPSOUT computes the integrated $ep$ and $\gamma^* p$
cross sections according to expression~(\ref{sigma}) and closes
histogram and n-tuple files.

\subsection{Implementation and usage}

DIPSI is written in standard FORTRAN 77 and uses the PATCHY offline
editor~\cite{patchy} to allow easier modifications. All routines are
contained in a PAMfile; a short
``cradle'' program contains the list of the routines that the user wants
to include and the modifications to the code, if any, that need
to be applied. By running PATCHY on the PAMfile a FORTRAN program
is produced.

The CERN
libraries PACKLIB and KERNLIB~\cite{cernlib} are necessary; 
extensive use is made of
the format free reading package FFREAD \cite{ffread} and of
the histogramming package HBOOK~\cite{hbook} (which is a part of
PACKLIB). The parton distribution package PDFLIB~\cite{pdf} may be needed 
(cf. appendix~\ref{gluon}).

\subsubsection{Input}

All necessary input parameters are read in from the user control cards
(logical unit 8) described in appendices~\ref{uor} and~\ref{control_cards}. 
A second input file (logical unit 9) is used in order to keep track of
the random number seeds. These are the only input files.

Among the input quantities are the 
gluon distribution, the strong coupling constant and the proton form 
factor. Concerning these parameters a remark is in order.
The cross section factorises in
$\alpha_s^2$, in the gluon density squared and in the proton form factor
squared. It is thus not necessary to generate
new samples of events if one exists already and 
if any of these quantities is modified: it is
sufficient to reweight the events.
As an example, if events have been generated with
a gluon density $\bar x g(\bar x)=3(1-\bar x)^5$ and
one wants to change to a different distribution $\bar x g^\prime(\bar x,
\bar q^2)$, the weight of each event should be multiplied
by the factor $[(\bar x g^\prime)/(\bar x g)]^2$.
If desired, however, a user provided parametrisation can be 
included in subroutine USRGLS; in that case the switch USRGLU in 
the control cards should be set to 1. 
The default version of the subroutine contains, as an example, a call
to the MRSA parametrisation from
the library PDFLIB~\cite{pdf}; the user can replace it with any other.

\subsubsection{Output}

DIPSI produces the following output files:

\begin{enumerate}
\item DIPSI.TST (logical unit 18), a log-file listing the
values of the input parameters selected via the control cards, the
integrated $ep$ (or $\mu p$) and the integrated $\gamma^* p$ cross
sections;
\item DIPSI.OUT (logical unit 6), containing the same log-file 
along with the HBOOK histograms in ASCII format;
\item DIPSI.HROUT (logical unit 40), with the same histograms
readable with PAW~\cite{paw};
\item DIPSI.NTP (logical unit 41), containing the n-tuple;
\item DIPSI.ERR (logical unit 19), containing (self-explanatory) error 
messages, if any; 
\item DIPSI.UNO (logical unit 17), containing all kinematic variables
and momenta for a single event. The event number is chosen via the
card JEVE.
\end{enumerate}

\begin{figure}
\vspace{-1.0cm}
\begin{center}
\leavevmode
\hbox{%
\hspace*{-0.8cm}
\epsfxsize = 13.5cm
\epsffile{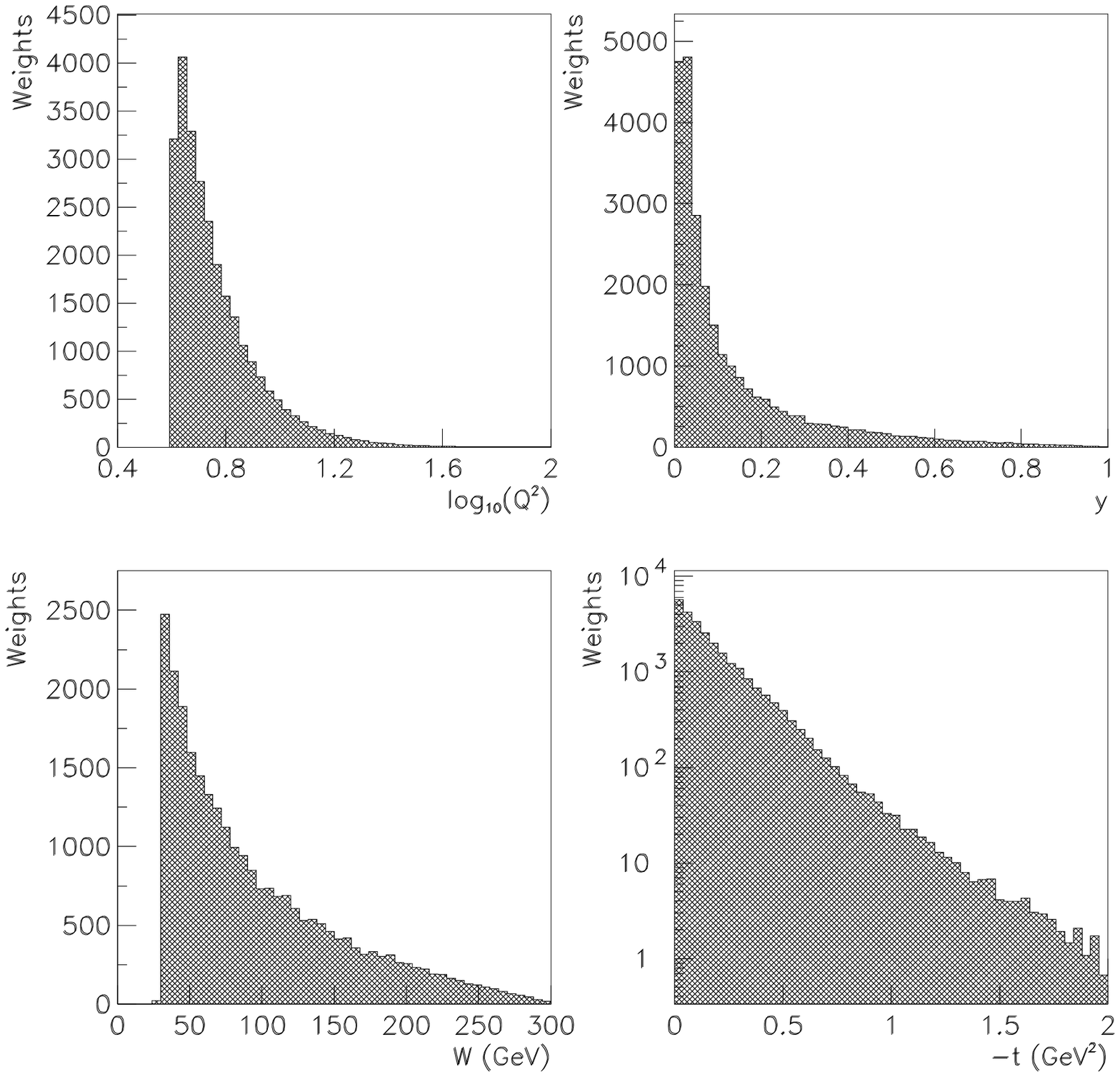}}
\end{center}
\vspace{-1cm}
\caption{Distributions over $Q^2$, $y$, $W$ and $t$ for events with elastic
production of $\rho^0$ mesons in the range $4<Q^2<100$~GeV$^2$
at HERA. See table~\protect\ref{generation_rho} for the list of the cards 
used.}
\label{results_rho}
\end{figure}

\begin{figure}
\vspace{-1.0cm}
\begin{center}
\leavevmode
\hbox{%
\hspace*{-0.8cm}
\epsfxsize = 13.5cm
\epsffile{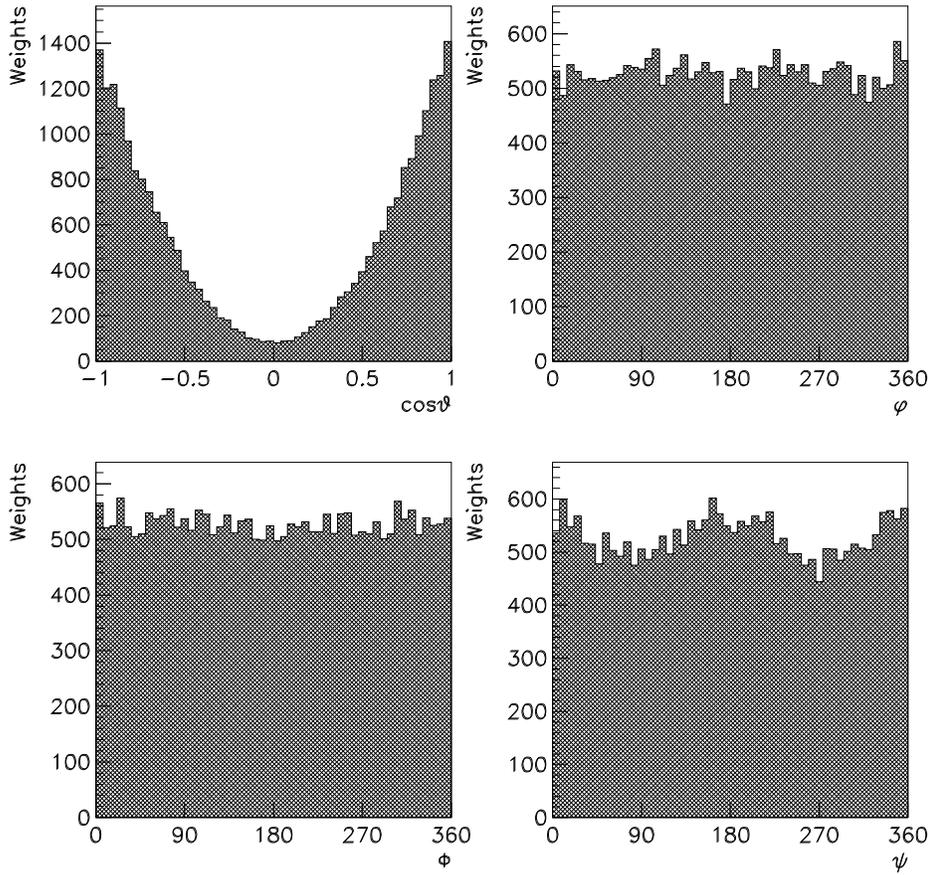}}
\end{center}
\vspace{-1cm}
\caption{Distributions over $\cos \vartheta$, $\varphi$, $\Phi$ 
and $\psi$ for events with elastic production of $\rho^0$ mesons in the 
range $4<Q^2<100$~GeV$^2$
at HERA. See table~\protect\ref{generation_rho} for the list of the cards used.}
\label{results_rho2}
\end{figure}

\begin{figure}
\vspace{-1.0cm}
\begin{center}
\leavevmode
\hbox{%
\hspace*{-0.8cm}
\epsfxsize = 13.5cm
\epsffile{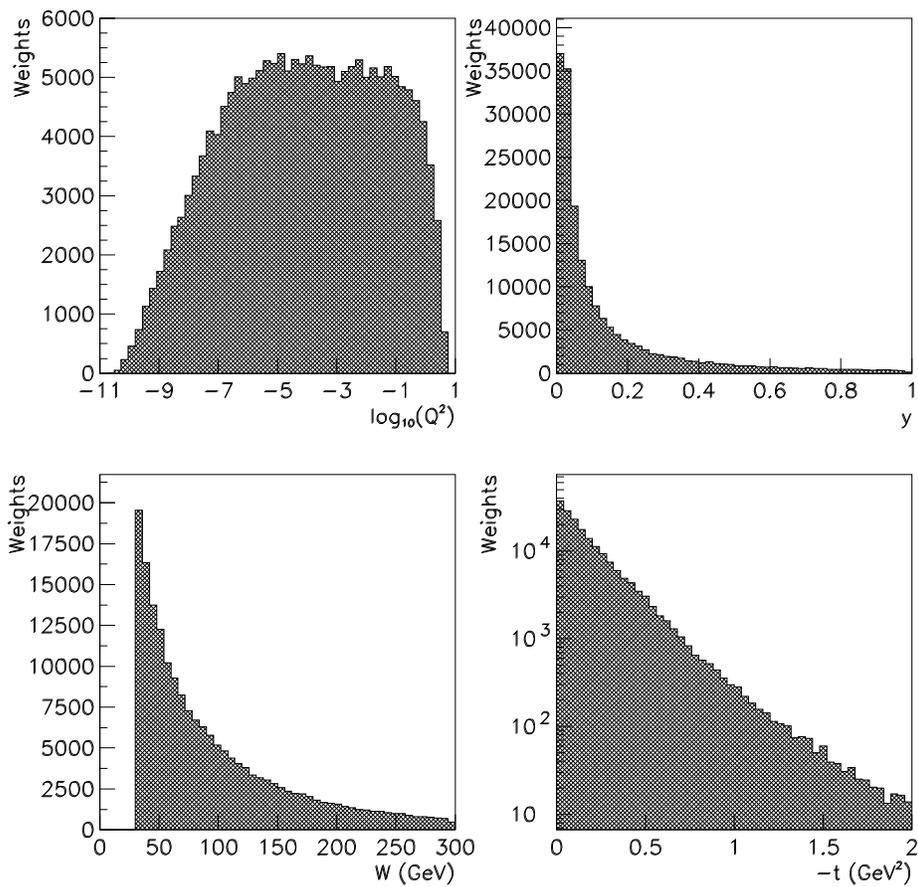}}
\end{center}
\vspace{-1cm}
\caption{Distributions over $Q^2$, $y$, $W$ and $t$
for events with elastic photoproduction of $J/\psi$ mesons at HERA. 
See table~\protect\ref{generation_psi} for the list of the cards used.}
\label{results_psi}
\end{figure}

\begin{figure}
\vspace{-1.0cm}
\begin{center}
\leavevmode
\hbox{%
\hspace*{-0.8cm}
\epsfxsize = 13.5cm
\epsffile{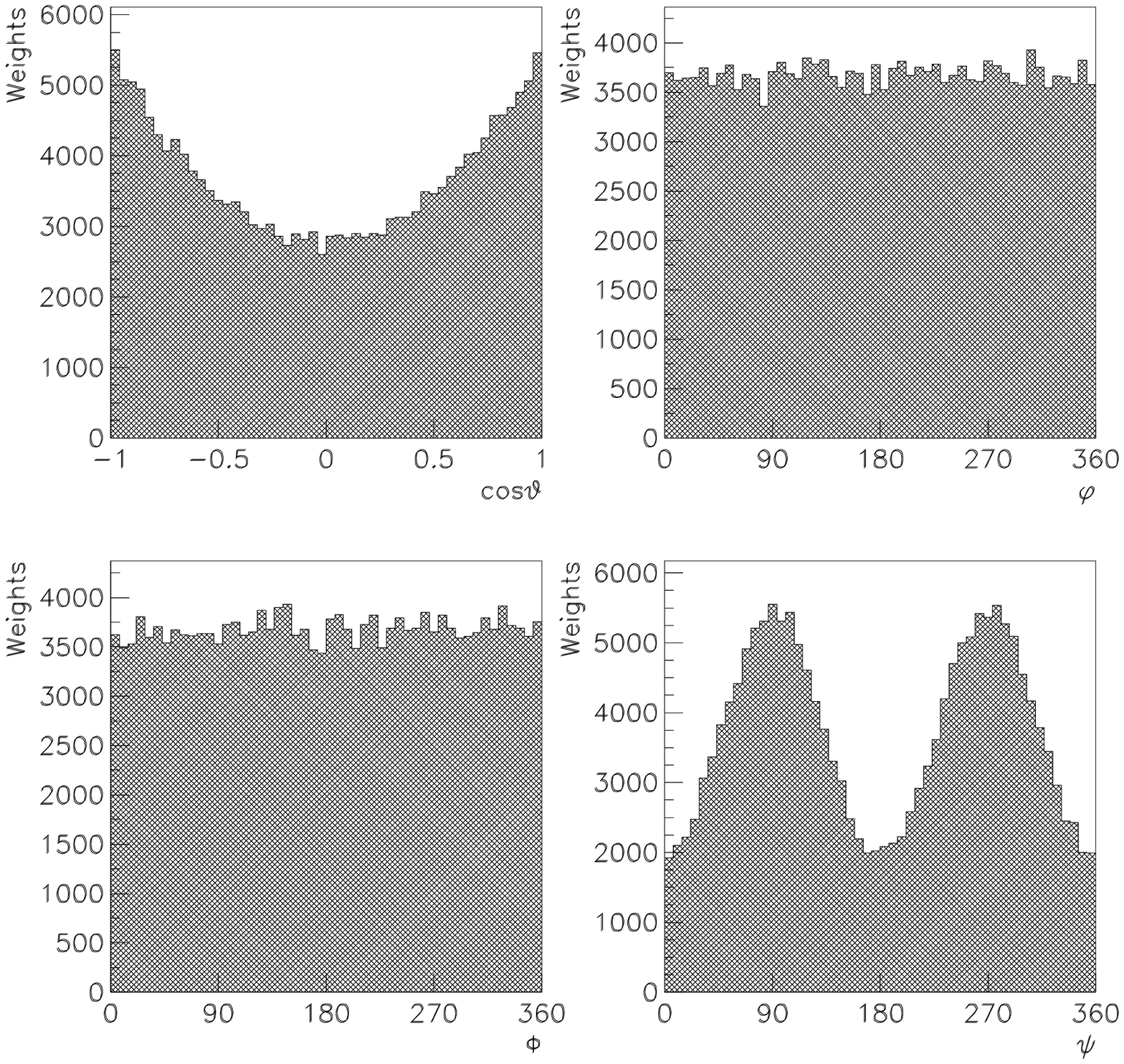}}
\end{center}
\vspace{-1cm}
\caption{Distributions over $\cos \vartheta$, $\varphi$, $\Phi$ 
and $\psi$ 
for events with elastic photoproduction of $J/\psi$ mesons at HERA. 
See table~\protect\ref{generation_psi} for the list of the cards used.}
\label{results_psi2}
\end{figure}

Figures \ref{results_rho} to \ref{results_psi2} show some typical 
plots. Figures \ref{results_rho} and \ref{results_rho2} show the
$Q^2$, $y$, $W$, $t$ and decay angles distributions for a sample
of $\rho^0$ mesons generated in the HERA kinematic regime, with 
$4<Q^2<100$~GeV$^2$ and $0.01<y<0.99$. Figures \ref{results_psi}
and \ref{results_psi2} show the same plots for a $J/\psi$ sample
produced in the photoproduction regime at HERA. For this sample 
$Q^2<4$~GeV$^2$ and $0.01<y<0.99$. Tables~\ref{generation_rho} and 
\ref{generation_psi} list the cards used to produce these samples
(see appendix~\ref{control_cards} for a description of the cards).

{\footnotesize
\begin{table}
\begin{center}
\begin{tabular}{|l|l||l|l|} \hline
EBEAM      & $-$27.5 & JDKLEP     &  2 \\
PBEAM      & 820.00  & IMASGE     &  0 \\
NTPFLAG    &     1   & MASMIN     &  0.3  \\
EMC        &     0   & MASMAX     &  1.5\\
NUTO       &  300000 & USRGLU     &  0\\
JEVE       &     0   & ICRXGX     &  0\\
YMIN       & 1.E$-$2 & IQ2EVO     &  0\\
YMAX       & 0.99    & IFORFA     &  0\\
YGEN       &   0     & FORFAS     &  2.5\\
QSQLOW     & 4.      & BIPT       &  3.0\\
QSQUP      & 100.      & PTMIN    & 0.0\\
KEWGEN     &      0  & PTMAX      & 10.000\\
JMESON     &      1  & ALPHAS     &  0.25\\
           &         & ETA        &  1.\\
\hline
\end{tabular}
\end{center}
\caption{Cards used to produce the sample of events with elastic
production of $\rho^0$ mesons in the range $4<Q^2<100$~GeV$^2$
at HERA, cf. figs.~\protect\ref{results_rho} and 
\protect\ref{results_rho2}. See appendix C for an explanation of the 
meaning of the cards. 
}
\label{generation_rho}
\end{table}
}

{\footnotesize
\begin{table}
\begin{center}
\begin{tabular}{|l|l||l|l|} \hline
EBEAM      & $-$27.5 & JDKLEP     &  0 \\
PBEAM      & 820.00  & IMASGE     &  0 \\
NTPFLAG    &     1   & MASMIN     &  0  \\
EMC        &     0   & MASMAX     &  0\\
NUTO       &  100000 & USRGLU     &  0\\
JEVE       &     0   & ICRXGX     &  0\\
YMIN       & 1.E$-$2 & IQ2EVO     &  0\\
YMAX       & 0.99    & IFORFA     &  0\\
YGEN       &   0     & FORFAS     &  2.50\\
QSQLOW     & 1.E-12  & BIPT       &  3.00\\
QSQUP      & 4.      & PTMIN      & 0.000\\
KEWGEN     &      0  & PTMAX      & 10.000\\
JMESON     &      0  & ALPHAS     &  0.25\\
           &         & ETA        &  1.\\
\hline
\end{tabular}
\end{center}
\caption{Cards used to produce the sample of events with elastic
photoproduction of $J/\psi$ mesons at HERA, 
cf. figs.~\protect\ref{results_psi} and \protect\ref{results_psi2}. 
See appendix C for an explanation of the 
meaning of the cards. 
}
\label{generation_psi}
\end{table}
}

\section{Summary}

We presented a Monte Carlo generator for elastic vector meson production
via reaction (\ref{reaction}), based on a leading logarithm QCD model
calculation \cite{misha}.

The model parameters are the strong coupling
constant, the two-gluon proton form factor and the gluon distribution
in the proton. The cross section is proportional to the square of the 
gluon momentum distribution in the nucleon.

The program correctly
describes the ZEUS data on photoproduction of $\rho^0$
\cite{zeus_rho,tesi_luc} and $J/\psi$ 
mesons \cite{zeus_psi}, 
and has been used to evaluate the ZEUS acceptance for these and other 
processes.
It is interesting that, with suitable input for the gluon distribution,
DIPSI reproduces also the $\rho^0$ photoproduction data, which are
in a $\bar q^2$ region ($\bar q^2\approx 0.15$~GeV$^2$) below 
that of applicability of perturbative QCD.

DIPSI is steered by control cards by means of which the user can select,
among other features, the type and decay mode of the vector meson, the
kinematic regime, the range of the independent variables and the distribution 
from which they are sampled.

Each event is assigned a weight proportional to its cross section. It is
thus possible to generate the independent variables according to
distributions different from those expected by the model, thereby
enhancing, if necessary, the statistical significance of the sample in
regions where the cross section is small.

\section{Acknowledgements}

We are indebted to Cristiana Peroni who proposed 
that we write DIPSI. 
We are grateful to her and to Aldus Whitfield for a critical reading 
of the manuscript. 
We would also like to thank Alexander Proskuryakov, Andrzej Sandacz 
and Ada Solano for many 
useful discussions.  

Some routines of DIPSI are based on similar ones belonging to 
generators for inelastic $J/\psi$ production developed 
within the EMC and NMC collaborations at CERN; we are grateful 
to Terry Sloan, Maarten de Jong and Chiara Mariotti for many discussions
on those programs and on the EMC and NMC $J/\psi$ data.

Many thanks are also due to our ZEUS and NMC colleagues who have used DIPSI 
up to now, thus actively contributing to the debugging process.

Finally, one of us (M.R.) is grateful to INFN for financial 
support while in Italy. 
\appendix

\section{List of the files provided}
\label{files}

The following files are provided:

\begin{verbatim}
DIPSI24.CAR
DIPSI.CRA
DIPSI.CARDS
DIPSI.COM
DIPSI.UNIX.
\end{verbatim}

They are respectively the PATCHY pamfile (card version) 
and cradle, the control 
cards, and the command-files in VMS and in UNIX.
The cradle file includes a PATCHY switch to be set depending on
whether the program is run in a UNIX or a VMS environment.

\section{Selection of model parameters}
\label{uor}

\subsection{The gluon momentum density}
\label{gluon}

In order to compute the gluon distribution in the proton, two
subroutines are provided, GLUONS and USRGLS. 

In GLUONS a few distributions are implemented. The 
default one is $\bar xg(\bar x) = 3(1-\bar x)^5$;
also available is the parametrisation obtained by the NMC 
experiment~\cite{nmglu}. 

\begin{figure}
\vspace{-1.0cm}
\begin{center}
\leavevmode
\hbox{%
\hspace*{-0.8cm}
\epsfxsize = 13.5cm
\epsffile{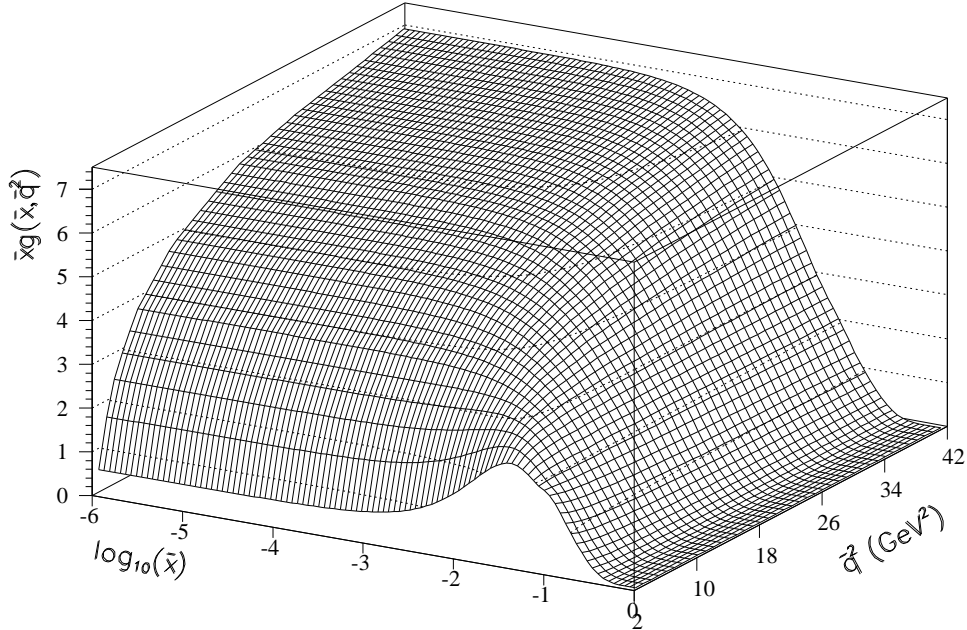}}
\end{center}
\vspace{-1cm}
\caption{The gluon 
distribution in the proton as parametrised in \protect\cite{antje}.} 
\label{evol}
\end{figure}

The card ICRXGX (cf. appendix~\ref{control_cards}) selects the available 
parametrisations:
\begin{itemize}
\item ICRXGX=0: $\bar xg(\bar x) = 3(1-\bar x)^5$;
\item ICRXGX=1: NMC lower limit for $\bar q^2=7$~GeV$^2$;
\item ICRXGX=2: NMC central value for $\bar q^2=7$~GeV$^2$;
\item ICRXGX=3: NMC upper limit for $\bar q^2=7$~GeV$^2$.
\end{itemize}
\noindent
Furthermore, if ICRXGX=2:
\begin{enumerate}

\item IQ2EVO=1: a $\bar q^2$ dependent parametrisation~\cite{antje} of the 
NMC results is used (fig.~\ref{evol}); 

\item IQ2EVO=2: same as IQ2EVO=1 but 
$\bar xg(\bar x, \bar q^2)=\bar xg(\bar x, \bar q^2_{\min})$ for $\bar q^2 
\le \bar q^2_{\min}= 2$~GeV$^2$;

\item IQ2EVO=3: same as IQ2EVO=2 but 
$\bar xg(\bar x, \bar q^2)= \bar x_{\min} g(\bar x_{\min}, \bar q^2)$ for 
$\bar x \le \bar x_{\min}= 3 \times 10^{-3}$.
\end{enumerate}
\noindent 

The routine USRGLS is a dummy routine with arguments $\bar x$,
$\bar q^2$ and $\bar xg$ which returns the
value of $\bar xg(\bar x,\bar q^2)$ for the given $\bar x$ and
$\bar q^2$. The user can implement the desired parametrisation here.
The default version of the subroutine contains, as an example, a call
to the MRSA parametrisation from
the library PDFLIB~\cite{pdf}.

Finally it is possible to switch from GLUONS to USRGLS by
means of the integer card USRGLU: USRGLU = 1 allows the
use of USRGLS (see appendix~\ref{control_cards}).

\subsection{The two-gluon form factor of the proton}

The choice of the proton form factor can be done directly via cards 
(cf. appendix~\ref{control_cards}).
The card IFORFA allows to choose between the dipole form factor,
$F_N(t) = 1/(1-\frac{t}{0.71})^2$,
(IFORFA=0) and an exponential one (IFOR\-FA=1). In the latter case the 
slope of the exponential can be set via card FORFAS.

\subsection{The strong coupling constant $\alpha_s$}

The value of the strong coupling constant $\alpha_s$ is determined 
by card ALPHAS (see appendix~\ref{control_cards}). A card value in the 
range between 0 and 1 fixes $\alpha_s$ to the card value. ALPHAS values 
outside the range 0-1 let $\alpha_s$ evolve with $\bar q^2$, assuming
$\alpha_s(\bar q^2)=12\pi/[25\ln{(\bar q^2/\Lambda_{QCD}^2)}]$; in 
this case, if $\alpha_s$ comes out to be larger than 0.7, 
it is automatically set to 0.7. The value of $\Lambda_{QCD}$
is set to 200~MeV.

\subsection{Relativistic effects in the meson wavefunctions}
\label{eta}

It is possible to shift the normalisation of the cross section 
to account for relativistic effects in the meson wavefunctions.
The cross section can be multiplied by a factor $[\eta^2]$; 
the user can choose it by setting the card ETA to the desired value. 
Zero or negative values are equivalent to setting $\eta=1$. 
For the $\rho^0$ meson a value $\eta=1.8$ is 
recommended~\cite{relativistic}. For $J/\psi$ mesons $\eta \approx 1$.

\section{Input control cards}
\label{control_cards}

An example of input control card file (logical unit 8),
with a description of each entry, is given below. For more details on the 
cards controlling the gluon distribution, the form factor of the 
proton, $\alpha_s$ and $\eta$, see appendix~\ref{uor}.

\begin{verbatim}
C *** Lepton beam momentum in GeV (must be negative...)
EBEAM      -27.5
C *** Proton beam momentum in GeV
PBEAM      820.00
C *** ntpflag = 0/1: do not fill/fill n-tuple
NTPFLAG         1
C *** emc = 0: electron beam, emc = 1: muon beam
EMC             0
C *** Total number of events to be generated 
NUTO           11000
C *** Event to be dumped in dipsi.uno
JEVE            4
C *** y generation range and spectrum: YGEN = 0 --> 1/Y, 
C ***                                  YGEN = 1 --> flat
YMIN        4.E-2
YMAX        9.E-2
YGEN            1
C *** Q2 generation range (in GeV2) and spectrum: 
C *** KEWGEN = 0 --> 1/Q2, KEWGEN = 1 --> 1/Q4, KEWGEN = 2 --> flat
QSQLOW     .10E-9
QSQUP     5.00E+0
KEWGEN          0
C *** JMESON: vector meson type
C             0  J/psi (also any integer other than the ones below)
C             1  rho
C             2  phi
C             3  Upsilon
C             4  omega
C             10  psi'(3686)
C             11  rho'(1450)
C             21  rho'(1700)
JMESON          0
C *** JDKLEP: decay mode
C             0  mu+ mu- (also any integer other than the ones below)
C             1  e+ e-
C             2  pi+ pi-
C             3  K+ K-
C             4  K0s K0l
C             5  pi+ pi- pi0 (with pi0 --> gamma gamma)
C             10  pi+ pi- psi --> pi+ pi- mu+ mu-
C             11  pi+ pi- psi --> pi+ pi- e+ e-
C             12  pi+ pi- rho --> pi+ pi- pi+ pi-
C             15  pi0 pi0 psi --> pi0 pi0 mu+ mu- (pi0s do not decay)
C             16  pi0 pi0 psi --> pi0 pi0 e+ e- (pi0s do not decay)
C             17  pi0 pi0 rho --> pi0 pi0 pi+ pi- (pi0s do not decay)
JDKLEP         1
C *** IMASGE: Vector meson mass distribution
C             0  Breit Wigner (also any integer other than the 
C                              ones below)
C             1  flat
C             2  Soeding (cf. ref. [28] and [10])
C             3  relativistic p-wave BW
IMASGE          0
C *** MASS GENERATION RANGE
C 
MASMIN       1.00
MASMAX       2.40
C *** Input gluon distribution (cf. appendix B.1)
USRGLU          0
ICRXGX          0
C *** Q2 evolution:    IQ2EVO=0 --> no q2 evolution,
C ***                  IQ2EVO=1 --> NMC xg(x,q2)
C *** (cf. appendix B.1)
IQ2EVO          0
C *** Proton form factor: IFORFA=0 --> dipole, 
C ***                     IFORFA=1 --> exponential 
C *** (cf. appendix B.2)
IFORFA          0
C *** Form factor slope (cf. appendix B.2)
FORFAS       2.50
C *** Pt2 (exponential) generation spectrum and range (in GeV2) 
C *** BIPT = generation slope (in GeV-2)
PTMIN       0.000
PTMAX    1000.000
BIPT         5.00
C *** Alpha strong (cf. appendix B.3)
ALPHAS       0.25
C *** normalisation shift to account for 
C *** relativistic wavefunction (cf. appendix B.4)
ETA          1.
C
STOP
\end{verbatim}

\section{Contents of the n-tuple}
\label{ntple}

Logical unit 41 is an unformatted file in which we store
many of the relevant kinematic variables as an n-tuple. In the 
following we give a brief description of the contents of 
this n-tuple. For a definition of the symbols the reader is referred to 
tables \ref{variables1} and \ref{variables2}. The n-tuple ID number is 666.

The components of particle momenta are given in the following
order: $x$, $y$ and $z$ components of three-momentum, energy 
and mass. The $z$ axis coincides with the incoming proton direction and 
is opposite to the direction of the incoming electron. If the proton is 
at rest the $z$ axis is also taken to be opposite to the direction of 
the incoming electron. Energy, momenta and masses are expressed in GeV;
cross sections are in nb.
Lines from 56 to 70 are relevant only for three-body decays. 
For these decays $\vartheta$ and $\varphi$ 
are the polar and azimuthal angles of the normal to the decay 
plane in the helicity frame. The term `unstable meson' refers to 
$\pi^0$, $\rho^0$ and $J/\psi$.

\noindent
\verb#***************************************************** # \\
\noindent
\verb#* NTUPLE ID=  666  ENTRIES= 100000   DIFFRACTIVE JPSI EVENTS # \\
\noindent
\verb#***************************************************** # \\
\noindent
\verb#* Var numb *   Name   *    Lower     *    Upper     * # \\
\noindent
\verb#***************************************************** # \\
\noindent
\verb#*     1    * Q2       * 0.100249E-09 * 0.499924E+01 * #$Q^2$\\
\noindent
\verb#*     2    * Y        * 0.100000E-01 * 0.499995E+00 * #$y$\\
\noindent
\verb#*     3    * NU       * 0.480671E+03 * 0.240333E+05 * #$\nu$\\
\noindent 
\verb#*     4    * PT2CM    * 0.194709E-05 * 0.466549E+01 * #$p_t^2$ \\
\noindent
\verb#*     5    * WSQ      * 0.899670E+03 * 0.451005E+05 * #$W^2$\\
\noindent
\verb#*     6    * Z        * 0.997422E+00 * 0.100000E+01 * #$z$\\
\noindent
\verb#*     7    * T        * -.600013E-03 * 0.467034E+01 * #$-t$\\
\noindent
\verb#*     8    * XL       * 0.979781E+00 * 0.999787E+00 * #$x_l$\\
\noindent
\verb#*     9    * PT       * 0.139538E-02 * 0.215999E+01 * #$p_t^p$\\
\noindent
\verb#*    10    * XBAR     * 0.212857E-03 * 0.168337E-01 * #$\bar x$\\
\noindent
\verb#*    11    * Q2BAR    * 0.239770E+01 * 0.423783E+01 * #$\bar q^2$\\
\noindent
\verb#*    12    * HCOSTH   * -.999990E+00 * 0.999992E+00 * #$\cos \vartheta$\\
\noindent
\verb#*    13    * HPHI     * 0.123719E-02 * 0.628306E+01 * #$\varphi$\\
\noindent
\verb#*    14    * HPHIC    * 0.184483E-02 * 0.628304E+01 * #$\Phi$\\
\noindent
\verb#*    15    * HPSI     * 0.150895E-02 * 0.628255E+01 * #$\psi$\\
\noindent
\verb#*    16    * WEIGHT   * 0.897848E-05 * 0.546283E+01 * #$ep$ weight\\
\noindent
\verb#*    17    * WTGAMP   * 0.565991E-04 * 0.263028E+02 * #$\gamma p$ 
weight\\
\noindent
\verb#*    18    * EBE1     * 0.000000E+00 * 0.000000E+00 * #incoming\\
\noindent
\verb#*    19    * EBE2     * 0.000000E+00 * 0.000000E+00 * #lepton\\
\noindent
\verb#*    20    * EBE3     * -.275000E+02 * -.275000E+02 * #\\
\noindent
\verb#*    21    * EBE4     * 0.275000E+02 * 0.275000E+02 * #\\
\noindent
\verb#*    22    * EBP1     * 0.000000E+00 * 0.000000E+00 * #target\\
\noindent
\verb#*    23    * EBP2     * 0.000000E+00 * 0.000000E+00 * #proton\\
\noindent
\verb#*    24    * EBP3     * 0.820000E+03 * 0.820000E+03 * #\\
\noindent
\verb#*    25    * EBP4     * 0.820001E+03 * 0.820001E+03 * #\\
\noindent
\verb#*    26    * ESE1     * -.221037E+01 * 0.219890E+01 * #scattered\\
\noindent
\verb#*    27    * ESE2     * -.220185E+01 * 0.219534E+01 * #lepton\\
\noindent
\verb#*    28    * ESE3     * -.272250E+02 * -.137196E+02 * #\\
\noindent
\verb#*    29    * ESE4     * 0.137501E+02 * 0.272669E+02 * #\\
\noindent
\verb#*    30    * ESE5     * 0.511000E-03 * 0.511000E-03 * #\\
\noindent
\verb#*    31    * ESP1     * -.179924E+01 * 0.161310E+01 * #recoiling\\
\noindent
\verb#*    32    * ESP2     * -.197851E+01 * 0.187068E+01 * #proton\\
\noindent
\verb#*    33    * ESP3     * 0.803420E+03 * 0.819825E+03 * #\\
\noindent
\verb#*    34    * ESP4     * 0.803421E+03 * 0.819826E+03 * #\\
\noindent
\verb#*    35    * ESP5     * 0.938272E+00 * 0.938272E+00 * #\\
\noindent
\verb#*    36    * GAM1     * -.219890E+01 * 0.221037E+01 * #virtual\\
\noindent
\verb#*    37    * GAM2     * -.219534E+01 * 0.220185E+01 * #photon\\
\noindent
\verb#*    38    * GAM3     * -.137804E+02 * -.275000E+00 * #\\
\noindent
\verb#*    39    * GAM4     * 0.233082E+00 * 0.137499E+02 * #\\
\noindent
\verb#*    40    * GAM5     * -.223590E+01 * 0.428417E-02 * #\\
\noindent
\verb#*    41    * VEC1     * -.306449E+01 * 0.294946E+01 * #vector\\
\noindent
\verb#*    42    * VEC2     * -.298695E+01 * 0.297885E+01 * #meson\\
\noindent
\verb#*    43    * VEC3     * -.135730E+02 * 0.162611E+02 * #\\
\noindent
\verb#*    44    * VEC4     * 0.309709E+01 * 0.168128E+02 * #\\
\noindent
\verb#*    45    * VEC5     * 0.309690E+01 * 0.309690E+01 * #\\
\noindent
\verb#*    46    * MUP1     * -.322182E+01 * 0.326902E+01 * #first\\
\noindent
\verb#*    47    * MUP2     * -.339305E+01 * 0.353246E+01 * #decay\\
\noindent
\verb#*    48    * MUP3     * -.136807E+02 * 0.127291E+02 * #particle\\
\noindent
\verb#*    49    * MUP4     * 0.182204E+00 * 0.136921E+02 * #\\
\noindent
\verb#*    50    * MUP5     * 0.511000E-03 * 0.511000E-03 * #\\
\noindent
\verb#*    51    * MUM1     * -.303850E+01 * 0.310788E+01 * #second\\
\noindent
\verb#*    52    * MUM2     * -.338481E+01 * 0.358694E+01 * #decay\\
\noindent
\verb#*    53    * MUM3     * -.135949E+02 * 0.127755E+02 * #particle\\
\noindent
\verb#*    54    * MUM4     * 0.188318E+00 * 0.136271E+02 * #\\
\noindent
\verb#*    55    * MUM5     * 0.511000E-03 * 0.511000E-03 * #\\
\noindent
\verb#*    56    * PIZ1     * 0.000000E+00 * 0.000000E+00 * #third decay\\
\noindent
\verb#*    57    * PIZ2     * 0.000000E+00 * 0.000000E+00 * #particle\\
\noindent
\verb#*    58    * PIZ3     * 0.000000E+00 * 0.000000E+00 * #(``unstable"\\
\noindent
\verb#*    59    * PIZ4     * 0.000000E+00 * 0.000000E+00 * #meson: $\pi^0$, $\rho^0$, \\
\noindent
\verb#*    60    * PIZ5     * 0.000000E+00 * 0.000000E+00 * #$J/\psi$)\\
\noindent
\verb#*    61    * PH1      * 0.000000E+00 * 0.000000E+00 * #first\\
\noindent
\verb#*    62    * PH2      * 0.000000E+00 * 0.000000E+00 * #particle from\\
\noindent
\verb#*    63    * PH3      * 0.000000E+00 * 0.000000E+00 * #``unstable"\\
\noindent
\verb#*    64    * PH4      * 0.000000E+00 * 0.000000E+00 * #meson\\
\noindent
\verb#*    65    * PH5      * 0.000000E+00 * 0.000000E+00 * #\\
\noindent
\verb#*    66    * PHO1     * 0.000000E+00 * 0.000000E+00 * #second\\
\noindent
\verb#*    67    * PHO2     * 0.000000E+00 * 0.000000E+00 * #particle from\\
\noindent
\verb#*    68    * PHO3     * 0.000000E+00 * 0.000000E+00 * #``unstable"\\
\noindent
\verb#*    69    * PHO4     * 0.000000E+00 * 0.000000E+00 * #meson\\
\noindent
\verb#*    70    * PHO5     * 0.000000E+00 * 0.000000E+00 * #\\
\noindent
\verb#***************************************************** #

\section{List of routines and functions}
\label{lur}

\begin{verbatim}


BREIT      non-relativistic Breit-Wigner mass distribution
DCROSS     vector product between two vectors
DEPHI      azimuthal angle determination
DIPSI      main program
DIPSIGEN   generation of kinematic variables 
DIPSINI    initialisation
DIPSOUT    cross section evaluation and output
DPHELI     helicity angles determination
GLUONS     evaluation of gluon distribution 
JDK        two-body decay
JPRYSKIN   cross section determination
HELOMEGA   helicity angles for three-body decays
LORENZ     three dimensional Lorentz boost
LUBEND     closing of unformatted file (lun 40)
LUBOOK     histogram and n-tuple booking
LUFIL      histogram and n-tuple filling
NTPINIT    definition of n-tuple variables
RUOTA      reference system rotation
SODING     Soeding shape distribution
SUB5       subtraction of two vectors
TIMTEM     time usage evaluation 
TREDK      three-body decay (including decay of most unstable meson)
ULANGL     angle determination
USRGLS     user routine to evaluate the gluon distribution 
WIGNER     relativistic Breit-Wigner mass distribution

\end{verbatim}

\end{document}